\documentclass[]{article}
\usepackage[T1]{fontenc}
\usepackage[utf8]{inputenc}
\usepackage[english]{babel}

\usepackage{booktabs} %

\usepackage[]{todonotes}
\usepackage{hyperref}
\usepackage{graphicx}
\usepackage{listings}
\usepackage{xcolor}
\usepackage{csvsimple}
\usepackage{geometry}
\usepackage{longtable}
\usepackage{tabularx}
\usepackage{multirow}
\usepackage{lscape}
\usepackage[tikz]{mdframed}
\usepackage{placeins}
\usepackage{float}
\usepackage{longtable}
\usepackage{xltabular}
\usepackage{ltablex}
\usepackage{amssymb,amsfonts}

\newcommand{\clarity}[0]{\color{magenta}\textbf{(CLA)}\color{black}}
\newcommand{\completeness}[0]{\color{magenta}\textbf{(COM)}\color{black}}
\newcommand{\uniformity}[0]{\color{magenta}\textbf{(UNI)}\color{black}}
\newcommand{\timelyupdates}[0]{\color{magenta}\textbf{(UPN)}\color{black}}
\newcommand{\nondupefforts}[0]{\color{magenta}\textbf{(NDE)}\color{black}}

\newcommand{\qualitygate}[0]{\textcolor{teal}{\textbf{(QG)}}}
\newcommand{\formalism}[0]{\textcolor{teal}{\textbf{(F)}}}
\newcommand{\doctracking}[0]{\textcolor{teal}{\textbf{(DT)}}}
\newcommand{\macro}[0]{\textcolor{teal}{\textbf{(M)}}}

\mdfsetup{roundcorner=10pt,middlelinewidth=1.5pt,innerleftmargin=0.1cm,innerrightmargin=0.1cm}

\usepackage{authblk}

\begin{document}

\title{Documentation-as-code for Interface Control Document Management in Systems of Systems: a Technical Action Research Study}

\author[1,2]{H\'ector Cadavid\thanks{h.f.cadavid.rengifo@rug.nl}}
\author[1]{Vasilios Andrikopoulos\thanks{v.andrikopoulos@rug.nl}}
\author[1]{Paris Avgeriou\thanks{p.avgeriou@rug.nl}}
\affil[1]{University of Groningen, Groningen, The Netherlands}
\affil[2]{Escuela Colombiana de Ingenier\'ia, Bogot\'a, Colombia}

\date{}

\providecommand{\keywords}[1]
{
	\small	
	\textbf{\textit{Keywords---}} #1
}
%

\maketitle              %
\begin{abstract}
	The architecting of Systems of Systems (SoS), that is, of systems that emerge from the cooperation of multiple independent constituent systems, is a topic of increasing interest in both industry and academia. However, recent empirical studies revealed what seems to be an overlooked aspect of the architecting of SoS that is linked to major integration and operational issues: the interplay between the various disciplines involved in such an architecting process. This aspect becomes particularly relevant for the management of the interfaces between the SoS constituents, where such disciplines inevitably meet. In this paper, we present the results of the first cycle of a Technical Action Research (TAR) study conducted in cooperation between the authors and a group of practitioners involved in the long-running architecting process of a large-scale radio astronomy SoS project. This TAR is aimed at exploring potential improvements of the document-centered interface management approach currently followed in this project by adopting elements of the \textit{documentation-as-code} philosophy, which is widely adopted in the domain of software systems. As a result, a working proof-of-concept of an ICD (Interface Control Document) management approach was developed by the researchers and evaluated by the practitioners. The  results of the study and the corresponding lessons learned are reported in this work.
	
\end{abstract}


\section{Introduction}

The concept of System of Systems (SoS) is used across application domains such as defense, automotive, energy, and health care to describe a family of independent systems that cooperate to provide capabilities that cannot be delivered by the individual systems~\cite{ISO21839}. The architecting process of this kind of systems is known for being challenging due to the operational and managerial independence of their constituents~\cite{maier1998architecting}, which arguably explains the significant volume of work in this area~\cite{cadavid2020architecting}. More recently, a particular aspect of the architecting process of this kind of systems has also been highlighted as posing additional challenges: the interplay between the different engineering disciplines involved in its development, such as systems engineering, software engineering, and electrical engineering, among others~\cite{sheard2018incose}. Studying this aspect is particularly important as recent empirical studies suggest that the interaction between those disciplines often leads to major integration and operational issues in SoS when their architecting practices are not harmonized~\cite{cadavidsurvey,cadavid2021system}. %
A specific pain point identified by these studies is that of \emph{interface management}.

Interface management in the context of SoS usually involves a formalized description of the interfaces between the constituent systems, i.e., the definition of what must be done to interact with each other, commonly referred to as Interface Control Documents (ICD) or Interface Design Documents (IDD). These documentation artifacts ensure compatibility between the said constituents, prevent integration errors, and improve the quality of the whole system~\cite{rahmani_managing_2011}. However, empirical studies in the field~\cite{cadavidsurvey,cadavid2021system} suggest that the details on these ICDs, defined by one of the involved parties, are incomplete or not sufficiently clear for the other parties to work with. This, in addition to the domain knowledge gap between these disciplines, seems to be a common cause of \textit{misunderstandings} and \textit{wrong assumptions} in the process~\cite{sheard2018incose,cadavidsurvey}.

Aiming to address this issue, in this paper we present the results of the first cycle of a technical action research (TAR) study~\cite{wieringa2014technical} designed to investigate an alternative ICD management approach. The study is conducted with a group of engineers from ASTRON\footnote{\url{https://www.astron.nl/}}, the Netherlands Institute for Radio Astronomy. This cooperative work between researchers and practitioners was motivated by the result of a previous exploratory case study on the LOFAR and LOFAR2.0, a long-running large-scale radio-telescope SoS~\cite{cadavid2021system}, where the aforementioned ICD-management issues  %
emerged. The \textit{documentation-as-code} or \textit{docs-as-code (DaC)} philosophy~\cite{gentle2017docs}, which encourages the creation and maintenance of technical documentation as rigorously as software, emerged as a common ground idea (between researchers and practitioners) to be explored as an improvement to the existing document-centered ICD management approach. The researchers had already identified that according to a significant amount of gray literature in the area\footnote{A collection of conferences, video Casts and articles from practitioners is available at \url{https://www.writethedocs.org/guide/docs-as-code/}}, the DaC philosophy has been used successfully to address issues similar to the ICD-management ones discussed above, albeit in the domain of Software Systems. ASTRON practitioners, on the other hand, had already been exploring an ad-hoc DaC approach that involves transcribing parts of the ICDs into a machine-readable documents. These documents would then be  used for automating some of the repetitive and error-prone tasks, like artifacts generation, during the development process. Thus, the idea of applying this approach at an ICD level appeared sound for both the practitioners and researchers. 

This study offers as a result four main contributions. First, it characterizes ICD management issues in SoS that are linked to misunderstandings and erroneous assumptions while implementing (parts of) an interface between (sub-)systems. Second, it explores, in collaboration with practitioners, a DaC-inspired solution that is alternative to traditional document-centered ICD management approaches for addressing the aforementioned issues. Third, it proposes a set of tools for implementing DaC pipelines based on the proposed approach. Fourth, it provides evidence of the merit of the proposed approach, together with points for its improvement.

The rest of this paper is structured as follows:  Section~\ref{sec:backg_rw} summarizes the background and the works related to this study. Section~\ref{sec:research_method} presents the study design, and Section~\ref{sec:results} answers the stated research questions. Section~\ref{sec:discussion} elaborates on the lessons learned in the process, while Section~\ref{sec:conclusions} concludes the study.

\section{Background and Related work}\label{sec:backg_rw}

\subsection{\textit{DaC} --- Documentation-as-code}\label{sec:dac_desc}

In recent years the software engineering community has been steadily shifting from traditional documentation approaches using conventional word processors, wikis, or other collaborative editing system to the DaC philosophy~\cite{gentle2017docs}. With DaC, where documentation is managed in the same way as source code in modern software projects, the community has been aiming at improving well-known issues such as outdated or unreliable technical documentation. Consequently, DaC implies the use of lightweight text-based markup languages (instead of proprietary formats and authoring tools) for the documents, so that they can be managed with modern (and proven) source code-oriented version control systems like Git and their related collaboration and automation tools. These automation possibilities mean, in turn, that a DaC documentation pipeline can ensure, to some extent, the quality of the documents by automatically validating or testing critical elements before their publication (e.g., the validity of code snippets within the documents, broken links, etc.) Furthermore, a DaC pipeline can also ensure uniformity and improve maintainability by using vendor-independent text-based specifications (e.g., for diagrams) that can be automatically transformed into visual elements that suit the organization's conventions.

Although the most common target of this documentation philosophy is software artifacts (e.g., APIs), it has also been adopted for higher-level documentation in industrial settings, showing to be useful to improve problems of missing or outdated documentation~\cite{thomchick2018improving} by reducing the complexity of documentation maintenance~\cite{ozerova2020comparison}. Examples include the documentation of products in government systems~\cite{lambourne_2017}, architecture documentation in transport systems\footnote{Deutsche Bahn - DB Systel - \url{https://github.com/docToolchain/docToolchain}}, and product engineering documentation\footnote{OpenGADES (a work in progress) - \url{https://wiki.eclipse.org/OpenADx}}. Moreover, despite this topic having received little attention by researchers~\cite{rong2020devdocops}, it is worth highlighting the large community of technical writers working on it, with the near 2000 members of the DaC-global network in the \textit{Write The Docs}\footnote{\url{https://www.writethedocs.org}} community as a prominent example.

\subsection{\textit{ICD management approaches}}\label{sec:related-w-icdm}

In the context of Systems Engineering, an ICD is a formal description of an agreement for the interfacing between two or more systems. There are no conventions nor standards to define these artifacts, as they usually differ from one company to another~\cite{rahmani_managing_2011}, even within the same application domain~\cite{louadah_towards_2014}. However, when it comes to the approaches to manage these artifacts, it is fair to say that the existing ones fall somewhere between the two ends of the spectrum~\cite{harvey2012document}: from pure model-centric, i.e., following a \textit{model-based systems engineering process} (MBSE), to pure document-centric ones. On the pure model-based end, the overall process is centered on a model about the system, from which documents like the ICDs are generated, when required, as a report. Examples in the context of SoS can be found in industries such as \textit{astronomy}~\cite{karban_verifying_2018,chiozzi_designing_2018}, \textit{space}~\cite{di_maio_interface_2018,vipavetz_interface_2016}, and \textit{defense}~\cite{tsui2018digital}. In these cases, with the exception of the defense industry, which has adopted the \textit{Unified Profile for DoDAF/MODAF} (UPDM), SysML is seemingly the de facto formalism in model-based approaches~\cite{japs2021save}. 

On the other extreme, the document-centric one, ICDs are mostly textual documents created with propietary word processing and diagramming tools, which evolve as interfaces are identified, defined, documented, and modified over time~\cite{wheatcraft20109}. This approach, despite the growing popularity of model-based ones as a response to its limitations, is still widely used in industry~\cite{broy2021advanced}. This is evidenced not only by examples in the literature in the \textit{aeronautics} industry~\cite{guo2020interface} and \textit{radio-astronomy} (the LOFAR case)~\cite{van2013lofar}, but in the many ICDs publicly available online for other domains\footnote{
	Particular examples are available in domains like \href{https://www.hhs.gov/guidance/sites/default/files/hhs-guidance-documents/DDC_ICD_V020117_041019_v1_5CR_040919_RETIRED_2.pdf}{\textit{\underline{Healthcare}}}, 
	\href{http://www.in2rail.eu/download.aspx?id=40938a15-24c4-427a-9e29-6811fabebaef}{\textit{\underline{Transport systems}}}, 
	\href{http://www.h2020-ergo.eu/wp-content/uploads/ERGO_D1_3_InterfaceControlDocument_V2.2.pdf}{\textit{\underline{Aerospace/robotics}}}, and
	\href{https://eoepca.github.io/master-system-icd/current/\#mainOIDC}{\textit{\underline{Astronomy}}}}. 
Between these two extremes, there are approaches that neither follow an MBSE process nor use text-based documents for the ICDs. Instead, these make use of computer aided tools to model only the relationships between the parties involved in the described interface (rather than the overall system), so that computations (e.g., evaluating the impact of a change of a parameter) can be performed. Examples in the literature include the use of spreadsheets to model ICDs in \textit{subsea production}~\cite{yasseri_interface_2019} and \textit{astronomy} systems~\cite{borrowman_can_2016}; and UML for modelling the interfaces of \textit{cyber-physical systems}~\cite{rahmani_managing_2011}. 

Unlike these approaches, the approach explored in this TAR could be described as one that combines the best of both worlds: the flexibility and expressiveness of text-based documents (something existing modelling languages lack in some cases~~\cite{zdravkovic2017challenges}), with the computability of text-based formalisms, within a workflow that ensures the quality and traceability of the overall documentation.

\section{Research method}\label{sec:research_method}\label{sec:rmethod}

\subsection{Problem formulation and research questions}
The goal of this study can be described following the \textit{design problem} structure proposed by Wieringa~\cite{wieringa2014technical}:%

\begin{mdframed}
	\begin{quote}
		\textbf{Improve} the ICD management practices in the context of large-scale SoS such as the ones developed by the ASTRON organization,\\
		\textbf{By} adapting the DaC philosophy to this context,\\
		\textbf{Such that} the managed ICDs lead to less occurrences of assumptions and misunderstandings while working with them,\\
		\textbf{In order to} reduce potential integration and operational issues caused by these occurrences.
	\end{quote}
\end{mdframed}

From this problem formulation, the following research questions are derived:

\begin{description}
	\item[RQ1] What are the issues with ICDs management that cause assumptions and misunderstandings when working with these documents in SoS?
	\item[RQ2] What are the features required for a DaC-based ICD management approach to address such issues?
	\item[RQ3] What is the design of an ICD management pipeline that provides such features?
	\item[RQ4] To what extent can the designed ICD management pipeline address the identified ICD management-related issues?
\end{description}

\subsection{Research method}\label{sec:research-method}

This study is based on the Technical Action Research \textit{(TAR)} method~\cite{wieringa2014technical}, as we want to investigate an experimental treatment (a DaC approach for ICDs management), to help a client with previously identified problems (ASTRON) and learn about its effects in practice. A TAR study involves five iterative phases: \textit{diagnosing}, \textit{action planning}, \textit{action taking}, \textit{evaluation}, and \textit{specifying learning}. In this sense, \textit{action} refers to a treatment to address the identified problems. The way each phase contributes to answering the proposed research questions, and the methods used by each are described below. We start by describing the context in which this method was applied.%

\paragraph{\textbf{Context}}

The research context is characterized by the document-centered management approach followed by ASTRON on the LOFAR and LOFAR2.0 projects over the past 10+ years~\cite{cadavid2021system}. Radio telescopes such as LOFAR consist of a large number of omnidirectional antennas concentrated in multiple, geographically distributed stations which, when working together, provide an effective collecting area equivalent to 300,000 square meters~\cite{beck2015future}. LOFAR2.0, on the other hand, is an ongoing expansion of the scientific and technical capabilities of LOFAR that relies on the lessons learned from its predecessor~\cite{juerges2021lofar2}.

\paragraph{\textbf{Diagnosing phase}}

This phase of TAR is focused on exploring and extending the current understanding of ICD documentation problems identified in~\cite{cadavid2021system}. Previous to this phase, the general idea of adopting DaC for ICD management was pitched to the potential participants. Subsequently, this phased used a \textit{virtual focus group} with the ASTRON practitioners; a focus groups is a qualitative research method to collect data on a given topic through a group interaction~\cite{kontio2008focus}. The discussion points for the focus group were derived from the relevant findings of the aforementioned previous study~\cite{cadavid2021system}. The actual focus group session was geared toward the collection of more context and details on ICD-management issues experienced in the LOFAR/LOFAR2.0 project in order to answer \textbf{RQ1}.

\paragraph{\textbf{Action planning}}

This phase is focused on identifying, discussing, and choosing solutions to improve the issues experienced by practitioners. As the \textit{DaC} philosophy was defined as a core element up-front in the project formulation, the discussed solutions were oriented towards how to apply such a philosophy in the context of ICD management in a hardware/software-intensive SoS like LOFAR. 
This phase is comprised of three steps, and its results, that is to say the features required by the DaC pipeline, were used to answer \textbf{RQ2}. First, a generic ICD template\footnote{Available in the online appendix - \url{https://figshare.com/s/88278ac89667496401ca}} was distilled from a curated set of existing ICDs, including those provided by practitioners. This template characterized the common elements included in an ICD, the kind of hardware elements described, and the way these are described in the documents. Second, the template was annotated collaboratively by one of the researchers and a group of practitioners (see Table~\ref{tab:team_dist}) to highlight specific symptoms of the issues identified in the diagnosing phase, e.g., particular elements of a given type of hardware elements that are often missed despite being key. Finally, researchers, in cooperation with one of the practitioners, defined the features that the DaC pipeline should provide in order to address the identified issues, using their particular issue symptoms to gauge their applicability in the process. 

\paragraph{\textbf{Action taking}}

Here the solutions selected to improve the identified issues are implemented. Therefore, in this phase the features identified in the previous phase were turned into an actual design that integrates existing DaC tools with any additional custom artifacts required by said features. This design is the answer to \textbf{RQ3}, and was implemented as a functional proof-of-concept that practitioners can try out and evaluate.

\paragraph{\textbf{Evaluation}}\label{sec:rm-evaluation}

In this phase, the effects of the action are captured trough different data collection methods~\cite{petersen2014action}. This phase addressed \textbf{RQ4} by exploring the efficacy and fitness for purpose of the proposed documentation management approach to address the identified issues during the diagnosing phase. To this end, we chose to perform a single case mechanism experiment~\cite{wieringa2014single} with the group of ASTRON engineers described in Table~\ref{tab:team_dist}. In this experiment\footnote{Available at \url{https://anonymous.4open.science/r/doc-as-code-single-case-experiment-BC61}}, practitioners were asked to carry out a number of ICD management activities using an online instance of the functional proof-of-concept created in the action taking phase. In the process, one of the researchers provided help with the process through Slack, so that observations or issues that emerged can also be included in the analysis to answer RQ3 and followed through in the consequent \textit{learning specification phase}. Once the practitioners completed the exercise, they filled up an online survey aimed at measuring the applicability of the proposed features in the identified ICD-management issues.

\paragraph{\textbf{Specifying learning}}

Finally, the lessons learned from this study (which is the first cycle of the action research), and their implications in future iterations of the study and the proposed documentation pipeline, were collected and are reported in the \textit{Discussion} section of the paper. %

\paragraph{\textbf{Participants and timeline}}
A total of ten people participated in this TAR study, which took place between May 2021 and the beginning of May 2022: two researchers and eight ASTRON practitioners. One of the researchers actively participated in the design and development activities during the \textit{action planning} and \textit{action taking} phases; both researchers took part in the research activities, designing the research instruments, and drawing lessons from the \textit{action evaluation} phase. The eight participants on the ASTRON side, on the other hand, were free to decide in which phase of the study they will participate. As a result, they participated in different phases of the study, as described in Table~\ref{tab:team_dist}.

\begin{table}[t]
	\small
	\begin{tabularx}{\columnwidth}{cp{0.4\textwidth}>{\centering}X>{\centering}X>{\centering}X>{\centering}X>{\centering\arraybackslash}X}
		\toprule
		\bfseries Participant & \bfseries Role in LOFAR project        & \bfseries D & \bfseries AP & \bfseries AT & \bfseries EV & \bfseries SL \\ \midrule
		P1          & Researcher                   & \checkmark          &                 &               &            &                     \\
		P2          & Software Engineer            & \checkmark          & \checkmark               &               & \checkmark          &                     \\
		P3          & RF Electronics Engineer      & \checkmark          &                 &               &          &  \checkmark                     \\
		P4          & Senior Software Engineer     & \checkmark          & \checkmark               &               & \checkmark          &                     \\
		P5          & Head of Software Development & \checkmark          &                 &               &            &                     \\
		P6          & Senior Software Engineer     &            &                 &               & \checkmark          &                     \\
		P7          & Software Engineer            &            & \checkmark               &               & \checkmark          &                     \\
		P8          & System Engineer              &            & \checkmark               &               & \checkmark          &                     \\
		\midrule
		R1          & N/A                          & \checkmark          & \checkmark               & \checkmark             & \checkmark          & \checkmark                   \\
		R2          & N/A                          & \checkmark          &                 &               & \checkmark          & \checkmark              \\     
		\hline
		          &    &   \scriptsize{May 2021- Jun 2021}  &   \scriptsize{Aug 2021- Jan 2022} & \scriptsize{Jan 2022- Mar 2022}   &  \scriptsize{Mar 2022- May 2022}   & \scriptsize{Apr 2022- May 2022}                   		
		\\ \bottomrule
	\end{tabularx}
	\caption{\label{tab:team_dist}Practitioners involved in the different phases of the study, and when such phases took place. \textit{P1, P2 ... P8}: practitioners on the ASTRON side. \textit{R1, R2}: external researchers. Phases: \textit{(D)} Diagnosing, \textit{(AP)} Action Planning, \textit{(AT)} Action Taking, \textit{(E)} Evaluation, \textit{(SL)} Specification of Learning }
\end{table}

\section{Results}\label{sec:results}
\subsection{RQ1: What are the issues with ICDs management that cause assumptions and misunderstandings when working with these documents in SoS?}

A wide variety of interfacing-related issues were identified by the virtual focus group conducted for the \textit{diagnosing} phase of the study, including some not explicitly related to the ICDs management itself, e.g., related to the way the interfaces were designed. As such issues are out of the the scope of this TAR and can be eventually explored in a separate study, the following list only includes the ones that have an actual influence to the \textit{action planning} phase:

\begin{description}
	\item[Clarity and  Cross-domain understandability \clarity:] an ICD often contains terminology that may be clear for the people involved in its original version, but years later it could be interpreted differently. This is also the case when multiple disciplines are involved in the process (e.g., hardware and software engineering), as in many cases there are similar terms between these disciplines with a different meaning.
	
	\item[Completeness \completeness:] Some critical details, particularly in the description of the hardware-side of the interfaces, are often omitted in the ICDs. This could lead to risky assumptions, or to error-prone informal information exchange. Additionally, time-behavioral and state-related aspects of the interfaces are rarely included in the ICDs. In particular, scenarios that lead to a failure state are important to include, although it is not always feasible to describe all of them on an ICD.
	
	\item[Uniformity \uniformity :] The lack of uniformity between ICDs, in the context of an SoS where many interfaces are usually involved, lead to confusion and misinterpretation.

	\item[Timely update notifications \timelyupdates:]  Changes in ICDs are not announced but rather discovered by people while working with them. 
	
	\item[Nonduplicated efforts \nondupefforts:] Interfacing-related information is duplicated across ICDs and the artifacts derived from it, e.g. intermediate ad hoc machine-readable formats and other tools created to support the development process. The time and effort required to keep in sync all these information sources is substantial and it could instead be invested in the actual engineering/development activities of the interfaces.
	
\end{description}

It should be noted that the lack of dynamic behavior details aspect of the completeness issue mentioned above, was not treated at this cycle of the study. As pointed out by the ASTRON part of the team, given the complexity this particular issue carries, addressing it would require a dedicated cycle within the TAR.

\subsection{RQ2: What are the features required for a DaC-based ICD management approach to address such issues?}

In the \textit{action planning} phase, four features were defined addressing the issues identified in the previous phase:

\begin{description}
	\item[Documentation-oriented quality gates \qualitygate] enforcing certain minimum quality criteria for the ICDs through centrally-managed rules. This involves automating the publication process of the ICDs through a CI/CD platform, with publication rules (i.e., quality gates) tailored to the context of technical documentation.	
	\item[Embedded machine-readable-formalisms \formalism] integrating existing formalisms for the description of technical elements within ICDs, particularly hardware-oriented ones. Such formalisms, when embedded to an ICD, can be automatically transformed into human-readable content (e.g., sections within the generated documents) and other artifacts (e.g., libraries). 
	\item[Document macros for automatic content generation \macro] providing custom extensions for the selected lightweight markup language used for ICD definition allowing engineers and technical writers to define where and how content will be automatically generated within an ICD.
	\item[Centralized document tracking \doctracking] keeping track of the publication status of the ICDs within the project/organization and of the dependencies between them. As this feature cannot be fully implemented through the version-control system (e.g., Git) itself, it entails creating a custom platform that would work in tandem with/on top of the CI/CD platform.

\end{description}

Table~\ref{tab:management-issues} describes particular symptoms of the issues identified in RQ1 and how the proposed features plan to address them. These symptoms indicate the occurrence of the corresponding issues in the context of actual ICDs (as described in Section~\ref{sec:research_method}) .

\begin{table}[t]
	\scriptsize
	\renewcommand{\arraystretch}{1.1}
	\begin{tabularx}{\columnwidth}{p{0.1\textwidth}p{0.35\textwidth}@{\hskip 0.1in}@{\hskip 0.1in}X}
	\toprule
	\bfseries Issue & \bfseries Symptom   &   \bfseries DaC feature \\
	\hline
	
	\multirow{6}{*}{\clarity} &  Overlapping/colliding abbreviations and terms.  & \macro~macros to (1) insert references to a centralized glossary, and (2) to generate a glossary section within the document accordingly. \qualitygate~for undefined abbreviations.\\
	
	\cmidrule{2-3}		
	
	 & Unclear language. & \qualitygate~number of violations to the writing style rules defined by the organization. \\
	
	\cmidrule{2-3}			
	
	 &  Broken links. & \qualitygate~Broken links metric. \\
	
	\hline		
		
	\multirow{6}{*}{\parbox{0.05\textwidth}{\completeness\\\uniformity}}&  Missing details: on specifications for reading/writing from/to a peripheral (e.g., endianness, R/W rights, update rates, etc.).;  on atomicity (e.g., which operations are atomic); on timing (e.g., commands timeout, ack signals timeout, etc.)  &  \formalism \macro~Support for hardware-oriented formalisms and \qualitygate~to validate the completeness of instances of such formalisms.\\
	
	\hline
	
	\multirow{13}{*}{\timelyupdates} &  ICDs should be amended when referenced documents are updated.  & \doctracking~Identify and notify when a document has a more recent timestamp than the ones that referenced it.\\		
		
		\cmidrule{2-3}
		
		& \multirow{9}{*}{\parbox{0.35\textwidth}{Changes are discovered rather than announced; outdated documentation.}}  & 	\formalism~Support for hardware-oriented formalisms  and \macro~macros for the generation of software artifacts (e.g., headers) and human-readable sections from the machine-readable formalisms. Based on this, and the \doctracking~platform, the artifacts'  checksum (e.g., headers)  previously generated by the documentation pipeline, and currently used in the development environment, could be automatically compared, during the (software) building process, against the most recent one (available online).\\
	
	\hline
		
	\multirow{13}{*}{\nondupefforts} &  Error-prone process of building a command's payload, (e.g., mapping parameter bits within a given index).  &
	\formalism~Support for hardware-oriented formalisms and \macro~macros for the automatic generation of software artifacts (e.g., library headers) and human-readable sections from the machine-readable formalisms.\\
	
	\cmidrule{2-3}

		&  Information replicated over multiple ICDs (no need to reiterate what is written elsewhere). & \doctracking~tracking of the multiple versions of an ICD (sources and published builds) and \macro~for referencing versions of other ICDs, reliably, through the centralized document tracking platform, and generating references section automatically.\\
		\cmidrule{2-3}

	    &  Need for the right balance between documentation maintenance efforts and actual engineering/development ones.  & 	\formalism~Support for hardware-oriented formalisms and \macro macros for the generation of software artifacts (e.g., headers) and human-readable sections from the machine-readable formalisms. \\
	
		\bottomrule
		
	\end{tabularx}
	\caption{ICD-management issues, their observed symptoms, and the identified associated features.}\label{tab:management-issues}

\end{table}

\begin{figure}[t]
	\centering
	\includegraphics[width=0.7\linewidth]{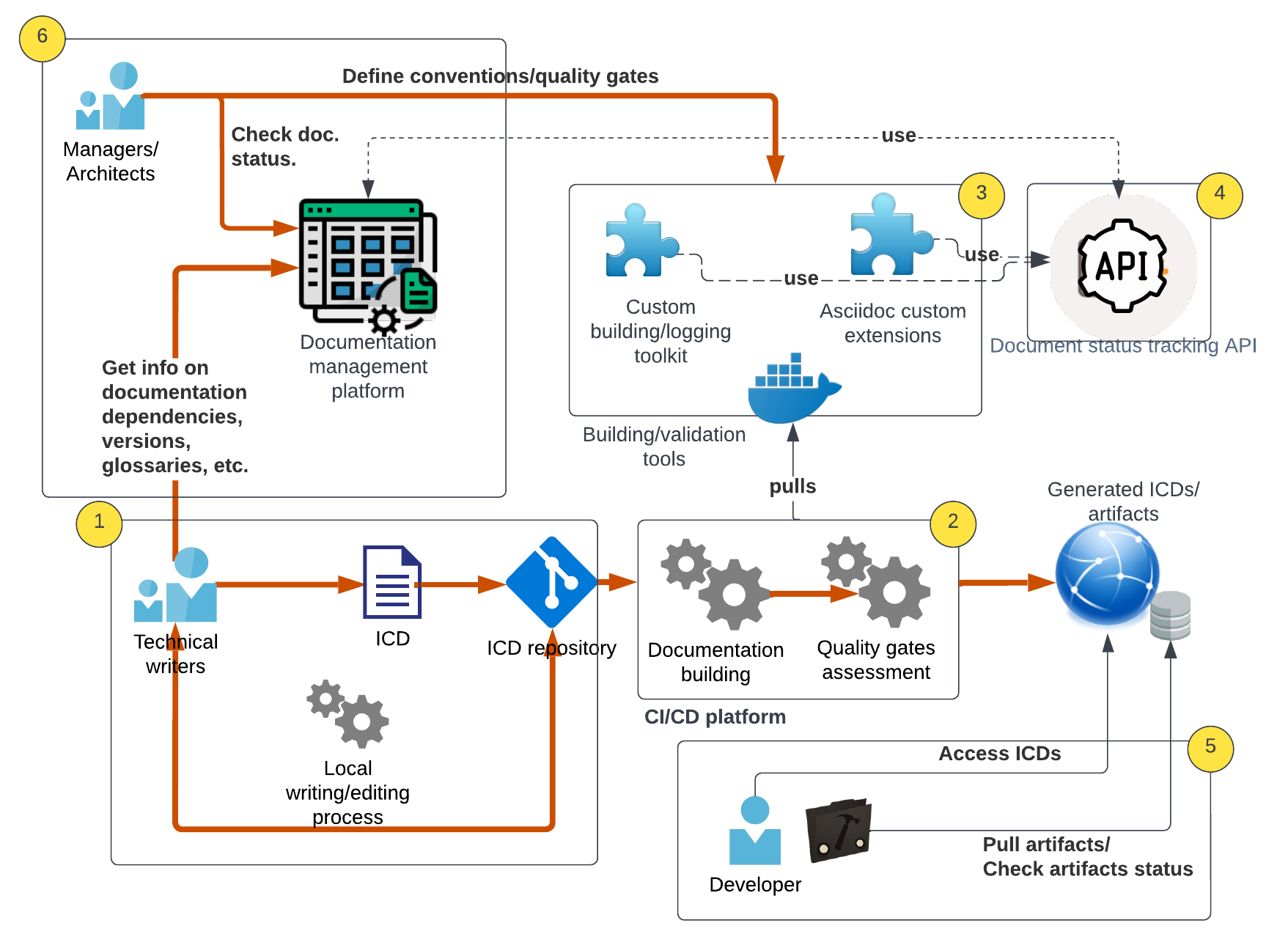}
	\caption{General overview of the ICD management pipeline}
	\label{fig:general-schema}
\end{figure}

\subsection{RQ3: What is the design of an ICD management pipeline that provides the identified features? }

In the \textit{action taking} phase, a solution was implemented as a functional proof of concept of a DaC pipeline instantiating the features identified in the previous phase.
The pipeline is depicted in Fig.~\ref{fig:general-schema}, and is described as follows. First, the writing/editing process of an ICD source document (1) is supported by any text editor since a text-based lightweight markup language is being used. 
For the proof of concept, Asciidoc was chosen as the lightweight markup language considering (in addition to its extensibility through custom macros) the variety of environments it can be interpreted on\footnote{For example, in front-end or backend, using
\href{https://github.com/asciidoctor/asciidoctor.js}{\textit{\underline{Asciidoctor.js}}} and \href{https://github.com/asciidoctor/asciidoctorj}{\textit{\underline{AsciidoctorJ}}} respectively.} and the flexibility this could provide to the solution on this, or follow-up cycles of the study.
The collaboration between writers is consequently mediated through a version management system like Git. 
Once a new ICD is ready for publication, the authors tag it as an official version. When such a tag is set, a CI/CD platform, linked to the ICD repository (2), launches the publication process using a centrally managed (at the project/organization level) building environment (3) that validates the minimum quality criteria in the process through  \textbf{Documentation-oriented quality gates} \qualitygate. This building environment, in turn, includes a number of custom extensions to enable the ICD-related \textbf{Document macros} \macro~and  in particular the ones that deal with the \textbf{Embedded machine-readable formalisms} \formalism. The events generated by the previous steps are reported to an API (4) as the means to have \textbf{Centralized document tracking} \doctracking~of the overall documentation status, and to provide context information to the building process of other documents (e.g., when dependencies between documents are involved). 

\begin{figure}[t]
	\centering
	\includegraphics[width=1\linewidth]{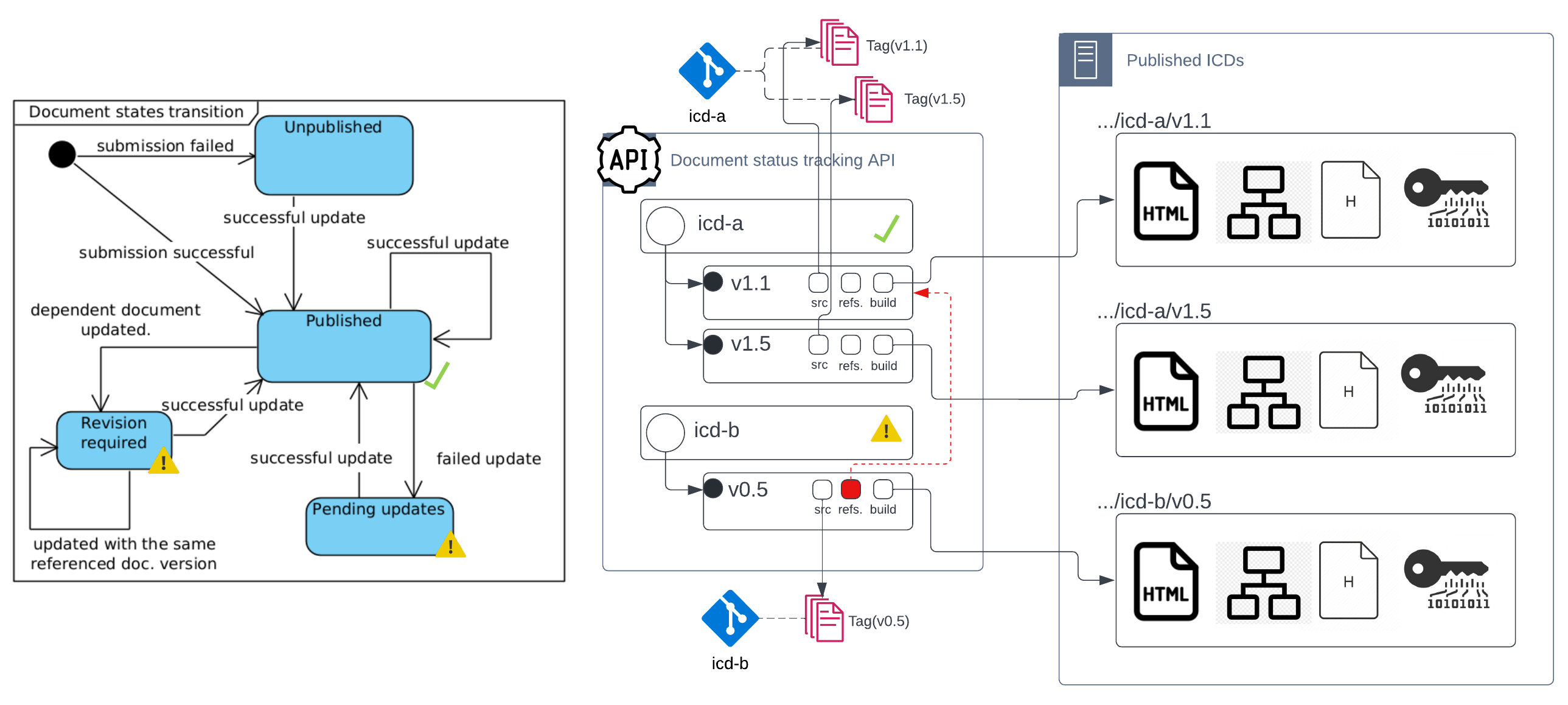}
	\caption{On the left, the diagram depicts the status transitions of the ICDs indexed in the centralized document tracking API. In the middle, the information indexed for each ICD is shown: (src) sources code, (refs) references to other ICDs, (build) canonical location of each version's generated document. Each published ICD (on the right), contains the generated documents, artifacts, and their corresponding unique checksum. In this particular case, document `icd-b' has a status of \textit{REVISION-REQUIRED}, as it is based on version 1.1 of `icd-a', which is no longer the most recent one.}
	\label{fig:central-tracking}
\end{figure}

Finally, the documentation and other generated artifacts are kept as the single source of truth for each interface by storing them in a canonical location (i.e., a URL accepted as the official one). With these locations the documentation users (e.g., software developers), besides accessing the latest version of an ICD, can integrate automated checks in their development environment (5) to verify that software artifacts are up-to-date with respect to the documentation they were generated from. 
The documentation management platform (6), on the other hand, has two purposes in the above process. On the one hand, for engineers or technical writers to have access to easy-to-read information about errors or failed quality gates during the ICD publication process. On the other hand, to serve as a dashboard that provides an overview of the overall documentation status.

In the following, the elements of the ICD-management pipeline that needed custom tools, tailored to suit the features identified in the previous section (RQ2), are described.

\vspace{-15pt}
\subsubsection{Quality-gates and Markup language extensions}

There are many tools that can perform a wide variety of quality checks in software source code within a CI/CD platform. However, when it comes to the source code of documentation, that is, markup language documents, quality criteria like the one defined in the \textit{action planning} phase (RQ2) require extending the interpreter of such a markup language so that the criteria are evaluated while the document is being processed.
These extensions (e.g., the \textit{custom macros} \macro) need not only to properly report their status during the building process (e.g., distinguishing between failed quality criteria and syntax errors) but also to work in tandem with the API ((4) in Fig.~\ref{fig:general-schema}). For this reason, a toolkit for building extensions with such features was created on top of the AsciidoctorJ\footnote{\url{https://docs.asciidoctor.org/asciidoctorj/latest/}} platform, and used to build the artifacts depicted in Fig.~\ref{fig:general-schema} (3). More details are reported in the project repository\footnote{\url{https://anonymous.4open.science/r/icd-dac-asciidoctorj-extensions}}.

\begin{figure}[t]
	\centering
	\includegraphics[width=0.65\linewidth]{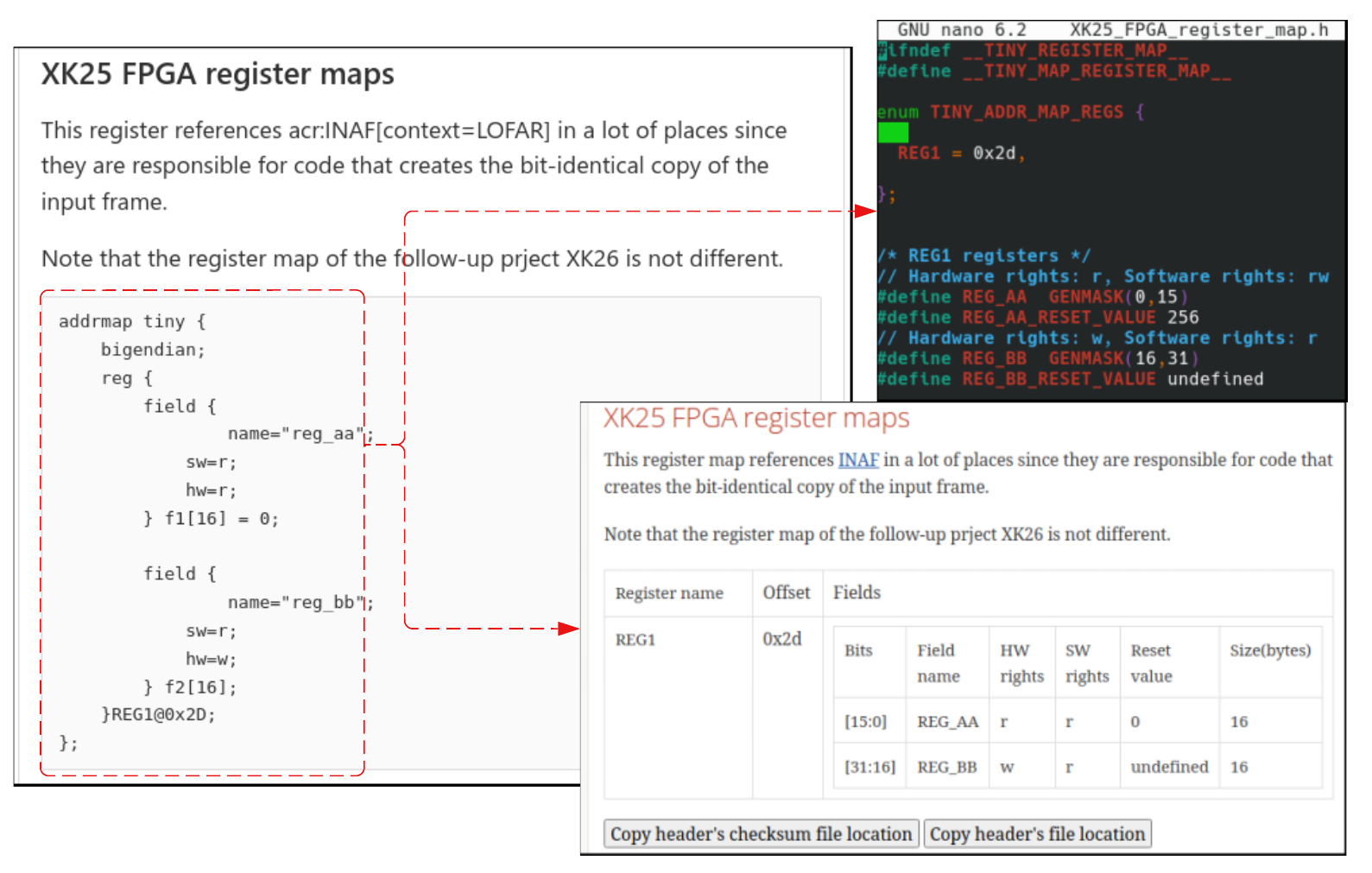}
	\caption{Excerpt from the source of a document created during an evaluation exercise (left), the generated source code headers (top-right), and the human-readable section expanded in the generated document (bottom-right).}
	\label{fig:sample-doc-transf}
\end{figure}
\vspace{-10pt}
\subsubsection{Centralized document tracking}

For the proposed DaC-based ICD pipeline, the generated documents are expected to become the single source of truth for all the development and testing tasks conducted on the interfaces described by it. Therefore, keeping track of the official documents and the sources from which they were generated is key to their proper evolution over time. On top of that, an ICD is often dependent on a specific version of other ICDs it is based on, and hence the validity or consistency of the former could be affected by changes in the latter. Given this, a documentation-status tracking API was implemented for the proposed documentation pipeline. This API, working in tandem with the aforementioned building components, keeps track of the locations of the documents and artifacts generated from each ICD, their dependencies, and their corresponding sources. Furthermore, it updates the status of the documents given the events captured by it as illustrated in Fig.~\ref{fig:central-tracking}, which, in turn, can be accessed through its front-end application ((6) in Fig.~\ref{fig:general-schema}). More details are reported in the project repository\footnote{\url{https://anonymous.4open.science/r/icd-management-platform-6D9A}}.

\vspace{-10pt}
\subsubsection{Macros for content generation \& embedded machine-readable formalism}

SystemRDL\footnote{\url{https://www.accellera.org/activities/working-groups/systemrdl}}, a widely adopted hardware description language was chosen to explore the use of a machine-readable formalism embedded in ICDs to describe hardware elements of interfaces \formalism. This involved the creation of a SystemRDL macro for Asciidoc that performs validations according to the defined completeness quality-gates, and generates a human-readable representation of the hardware along with base software artifacts (as depicted in Fig.~\ref{fig:sample-doc-transf}). With the automatic generation of check-sums of the said generated artifacts implemented in the extension, developers can check whether their local copy is up to date, as previously described.

\begin{table}[t]
	\scriptsize
	\begin{tabularx}{\columnwidth}{p{0.1\textwidth}X@{\hskip 0.1in}ccccc}
		\toprule
		& & \multicolumn{5}{c}{\bfseries Participants}\\
		\cmidrule{3-7}
		& \bfseries Question                  & P1 & P2 & P3 & P4 & P5 \\ \midrule
		Q1 \qualitygate & I believe that enforcing the expected minimum quality criteria of the texts within the ICDs, by means of automated ‘quality gates’, would improve their clarity and understandability. & A  & D  & SA  & SA  & N  \\
		Q2 \qualitygate & I believe that enforcing the expected minimum quality criteria of the technical details included on the ICDs (e.g., hardware-level details), by means of automated ‘quality gates’, would reduce the documentation-related integration/operational issues experienced in the past.     & A  & A  & SA  & SA  & N  \\ 
		Q3 \macro & I think that having part of the documentation (e.g., glossaries, hardware description, etc) and other artifacts (e.g., software components) generated from models embedded in the ICDs would reduce the documentation maintenance efforts when compared with the existing approach. & A  & N  & SA  & SA  & SA  \\ 
		Q4 \formalism & I think the automatic generation of human-readable content and software artifacts from formal models embedded in the ICDs (e.g., SystemRDL, ARGS, etc.), would improve the data-transcription-related issues experienced in the past.     & A  & A  & A  & SA  & SA  \\ 
		Q5 \doctracking & I think that having centrally managed ICDs, with their different versions/dependencies tracked, and status changes reported to relevant roles, would be helpful to mitigate the issues related to outdated documentation experienced in the past.     & A  & A  & A  & SA  & SA  \\ 				
		Q6 \doctracking & I think that integrating the information provided by the centrally-managed ICDs into the software development process, in a realistic setting, would be useful to avoid shipping software based on ICDs that could be outdated or that need to be revisited.     & A  & N  & A  & SA  & A  \\ 
		
		\bottomrule  
	\end{tabularx}
	\caption{Perceived applicability of the DaC features after using them in the proof-of-concept. Grades: (SA) Strongly Agree, (A) Agree, (N) Neigher agree nor disagree, (D) Disagree, (SD) Strongly disagree}\label{tab:surveyq}
\end{table}
Additional content generation extensions, based on the results of the \textit{action planning} phase of the study, include: automatic generation/validation of the document log section (based on Git history), references, and glossary. The source of a sample ICD written in Asciidoc, using the above extensions, and its corresponding output, is included in the online appendix\footnote{\url{https://figshare.com/s/88278ac89667496401ca}}.

\subsection{RQ4: To what extent can the designed ICD management pipeline improve the identified ICD management-related issues?}

The efficacy and fitness for purpose of the features identified in the \textit{action planning} phase and implemented in the \textit{action taking} phase was evaluated through an online survey filled by the ASTRON practitioners who participated in the \textit{evaluation} phase after conducting the \textit{single-case} mechanism experiment (see Section~\ref{sec:rm-evaluation}).  Table~\ref{tab:surveyq} shows the graded feedback collected by this survey. In the following we elaborate on these results based on the justification given by the participants as answers to open-ended question for their selected choices, and the Slack interaction in the process.

\subsubsection{Documentation-oriented quality gates}

The practitioners' perception on enforcing minimum quality criteria by means of quality gates were in general positive, as shown by questions Q1 and Q2 in Table~\ref{tab:surveyq}. However, the \textit{Disagreement} and \textit{Neutral} responses in Q1, about the enforcement of the style of the ICD elements described in natural language, were justified with doubts about its applicability in a more realistic setting. More specifically, two of the participants pointed out that (1) some text-based descriptions could be very hard to validate, if such validation is at a technical level, and (2) that writing styles are hard to impose and writers could find workarounds to avoid them.
The enforcement of technical elements of the ICDs like forgotten keywords and wrong ranges, on the other hand, was seen  in general as useful, although some interfacing parameters could be too difficult to validate automatically, especially the ones related to dynamic behavior, time dependencies and performance limitations. Overall, with the ICDs seen as the single source of truth and reference for engineers to work on an interface, the need for these automated quality gates would decrease in the long run, as will the time spent on resolving inconsistencies during development and testing.

\subsubsection{Document macros for automatic content generation}

According to the justifications for the grades given in Q3, having macros for content generation would help ICDs to become the single source of truth that can be referenced throughout the process. This would help, in turn, in identifying incorrect assumptions in the early stages of the process. On the other hand, the elements automatically generated through macros tailored for ICDs (content, artifacts, etc.), as long as they are reliable and reproducible, would not only improve their consistency but also free time to work on the content that cannot be automatically generated. However, although the above would improve the overall quality of the documentation, that would not necessarily be the case for its maintainability. In this regard, the analysis of Slack conversations during the evaluation exercise highlighted two areas of improvement. First, as a means to improve the workflow speed, an editing tool tailored to the proposed features, that is to say, that allows technical writers to identify errors or unfulfilled quality criteria before submitting changes to the documentation pipeline was deemed necessary. Second, the pipeline should support allowing locally-defined, document-specific glossary entries, in addition to the ones defined at  project/organization level.

\subsubsection{Embedded machine-readable formalisms}

Question Q4 (see Table~\ref{tab:surveyq}) and the corresponding open-ended responses also showed an overall positive perception on the applicability of an embedded formalism within the ICDs as a means to remove ambiguity through the uniformity and standardization of the artifacts automatically derived from it. This is seen as a must-have to prevent issues in different engineering groups. Furthermore, this feature was perceived as something that would save a considerable amount of time normally spent writing boilerplate ICD parts, and reduce the human factor that often causes issues when translating from ICD to source code.

\subsubsection{Centralized-documents tracking}

According to the practitioners' views on Q5 and Q6, a centralized platform makes sense if ICDs are expected to be the single source of truth within a project/organization. Moreover, adopting this systematic approach for managing ICDs implies that now it would be necessary to decide, at a higher level within the project/organization, when to go from one version to another across the whole product. Although this would add complexity to the overall process, it would be a positive change.
On the other hand, the generation of artifacts with version matching as supported by such a platform would prevent developers from using an outdated version when implementing parts of an interface. However, as pointed out by the participants, to properly close the loop failed version test results should be reported to the central ICD hub so that action can be taken. Furthermore, the tooling should also allow the interfaces to be round-trip converted from source-code back to the model.
On top of these observations, practitioners emphasized that although the documentation pipeline tooling itself would prevent small but annoying mistakes, most of the success of the interfacing process would still lie on the communication and cooperation between engineers. Moreover, it would also be important to ensure that engineering and maintenance teams also keep this documentation up-to-date during operations.

\section{Discussion/Lessons Learned}\label{sec:discussion}
The results suggest that the proposed approach, when applied to an actual ICD --- making it the \textit{single source of truth} for their related interfacing-related activities --- would indeed prevent wrong assumptions early in the process, ensure its uniformity, and improve its overall quality. Similar conclusions have been drawn for related approaches that make use of computer-aided tools (see Section~\ref{sec:related-w-icdm}) for this purpose. However, the proposed approach differs from these tools by allowing the specification of interfacing agreements (between the (sub-)systems that will be connected) combining a) the formality of models for describing critical, error-prone elements, with b) the flexibility of using natural language for all the other elements that such models could not describe. The proof-of-concept features related to the latter (i.e.~macros and quality gates for glossaries and writing style rules) were seen as nice to have in order to improve clarity and avoid misinterpretations. However, some practitioners' views suggest the need for further research on the applicability of writing style rules outside of the experimental evaluation setting. As a result, in future work we foresee replacing the generic writing rules currently used by the PoC with more tailored, domain-specific ones elicited from existing documentation and practitioners' input. 

The evaluation results also highlighted the importance of good user experience (UX), something that needs to be improved in the PoC, as the extra time imposed by the complexity of the proposed approach (when compared to a regular word processor) could exceed the time saved by its features (e.g., content generation). This includes the need for proper authoring tools and glossaries management that allows consistently combining and merging local glossaries with organization-level ones. As part of the work on the latter, collaboration with the \textit{shareable glossaries} project\footnote{\url{https://thegooddocsproject.dev/docs/glossaries/}} is envisioned in this direction.

The centralized-document tracking feature seems to enable the centralized management of dependencies between ICDs, which is an important element of the engineering process of SoS like LOFAR; such dependencies have neither been considered by previous ICD management approaches, nor supported by regular DaC tools. The results of the evaluation show that this feature makes sense as a means to ensure consistency and traceability between the said documents and the artifacts derived by them. In particular, the version matching mechanism that this feature enables, would ensure consistency by issuing warnings within the development environment. Moreover, closing the loop of this feature (as suggested after its evaluation), that is, reporting the failed version matches back to the central hub, would allow the organization to know when outdated references are being used and take timely actions. Overall, the proposed features would likely help the organization to decide when to move from one interface version to another, which (according to the practitioners) is a positive development.

Finally, it is worth highlighting, as pointed out by practitioners, that the proposed documentation management approach will not (and should not) remove the need for proper communication and collaboration while working on interfaces. Previous studies~\cite{cadavidsurvey,fairley2019systems} have confirmed that regardless of the tools or methodologies, proper collaboration and communication are still a key element of an engineering project success. However, by freeing the tension between the need to be flexible in terms of writing specifications and the need to create consistent and verifiable documentation, this approach has the potential to not only deliver high-quality ICDs over time, but also provide a solid foundation for such a collaboration.

\section{Conclusion}\label{sec:conclusions}
In this paper we report the results of the first cycle of a \textit{technical action research} study in which the authors and a group of ASTRON engineers explored \textit{documentation-as-code}, a philosophy of documentation management with a growing community of practice in software systems, to address ICD management issues identified in large-scale SoS. In doing so, a functional proof-of-concept of a documentation pipeline, with features tailored to particular symptoms of such issues in LOFAR, a long-running radio-astronomy SoS, was designed and implemented. A single-case experiment on ICD management scenarios exercised in said proof-of-concept, also with ASTRON practitioners,  was used for validation. According to the results, it is fair to say that this ICD management approach, which to our knowledge has not been explored before in this context, is promising to address some of the recurring issues with SoS of this type. Some areas for improvement are planned to be addressed in the following cycles of the TAR.

\bibliographystyle{splncs04}
\bibliography{references.bib}

\begin{thebibliography}{10}
\providecommand{\url}[1]{\texttt{#1}}
\providecommand{\urlprefix}{URL }
\providecommand{\doi}[1]{https://doi.org/#1}

\bibitem{beck2015future}
Beck, R.: Future observations of cosmic magnetic fields with lofar, ska and its
  precursors. In: Magnetic Fields in Diffuse Media, pp. 3--17. Springer (2015)

\bibitem{borrowman_can_2016}
Borrowman, A.J., Taylor, P.: Can your software engineer program your {PLC}? In:
  Software and {Cyberinfrastructure} for {Astronomy} {IV}. vol.~9913, p.
  99131S. International Society for Optics and Photonics (Jul 2016)

\bibitem{broy2021advanced}
Broy, M., B{\"o}hm, W., Rumpe, B.: Advanced systems engineering. In:
  Model-Based Engineering of Collaborative Embedded Systems, pp. 353--364.
  Springer (2021)

\bibitem{cadavid2020architecting}
Cadavid, H., Andrikopoulos, V., Avgeriou, P.: Architecting systems of systems:
  A tertiary study. Information and Software Technology  \textbf{118},  106202
  (2020)

\bibitem{cadavidsurvey}
Cadavid, H., Andrikopoulos, V., Avgeriou, P., Klein, J.: A survey on the
  interplay between software engineering and systems engineering during sos
  architecting. In: Proceedings of the 14th ACM / IEEE International Symposium
  on Empirical Software Engineering and Measurement (ESEM). ESEM '20,
  Association for Computing Machinery, New York, NY, USA (2020)

\bibitem{cadavid2021system}
Cadavid, H., Andrikopoulos, V., Avgeriou, P., Broekema, P.C.: System- and
  software-level architecting harmonization practices for systems-of-systems :
  An exploratory case study on a long-running large-scale scientific
  instrument. In: 2021 IEEE 18th International Conference on Software
  Architecture (ICSA). pp. 13--24 (2021)

\bibitem{chiozzi_designing_2018}
Chiozzi, G., Andolfato, L., Kiekebusch, M.J., Kornweibel, N., Schilling, M.,
  Zamparelli, M.: Designing and managing software interfaces for the {ELT}. In:
  Guzman, J.C., Ibsen, J. (eds.) Software and {Cyberinfrastructure} for
  {Astronomy} {V}. p.~78. SPIE, Austin, United States (Jul 2018)

\bibitem{di_maio_interface_2018}
Di~Maio, M., Atorf, L., Dahmen, U., Schluse, M., Rossmann, J., Hoppe, M.,
  Kapos, G.D.: Interface {Management} with {Closed}-{Loop} {Systems}
  {Engineering} ({CLOSE}). In: 2018 {IEEE} {International} {Systems}
  {Engineering} {Symposium} ({ISSE}). pp.~1--8 (Oct 2018)

\bibitem{fairley2019systems}
Fairley, R.E.: Systems Engineering of Software-enabled Systems. Wiley Online
  Library (2019)

\bibitem{gentle2017docs}
Gentle, A.: Docs like code. Lulu Press, Inc (2017)

\bibitem{guo2020interface}
Guo, D., Zhang, X., Zhang, J., Li, H.: An interface management approach for
  civil aircraft design. In: International Conference on Aerospace System
  Science and Engineering. pp. 435--446. Springer (2020)

\bibitem{van2013lofar}
van Haarlem, M.P., Wise, M.W., Gunst, A., Heald, G., McKean, J.P., Hessels,
  J.W., de~Bruyn, A.G., Nijboer, R., Swinbank, J., Fallows, R., et~al.: Lofar:
  The low-frequency array. Astronomy \& astrophysics  \textbf{556}, ~A2 (2013)

\bibitem{harvey2012document}
Harvey, D., Waite, M., Logan, P., Liddy, T.: Document the model, don’t model
  the document. In: Proc. Syst. Eng./Test Eval. Conf. 6th Asia Pac. Conf. Syst.
  Eng (2012)

\bibitem{ISO21839}
Draft bs iso/iec 21839 information technology - systems and software
  engineering - system of systems (sos) considerations in life cycle stages of
  a system. Standard, International Organization for Standardization, Geneva,
  CH (Mar 2018)

\bibitem{japs2021save}
Japs, S., Anacker, H., Dumitrescu, R.: Save: Security \& safety by model-based
  systems engineering on the example of automotive industry. Procedia CIRP
  \textbf{100},  187--192 (2021)

\bibitem{juerges2021lofar2}
Juerges, T., Mol, J., Snijder, T., et~al.: Lofar2. 0: Station control upgrade
  (2021)

\bibitem{karban_verifying_2018}
Karban, R., Troy, M., Brack, G.L., Dekens, F.G., Michaels, S.B., Herzig, S.:
  Verifying {Interfaces} and generating interface control documents for the
  alignment and phasing subsystem of the {Thirty} {Meter} {Telescope} from a
  system model in {SysML}. In: Angeli, G.Z., Dierickx, P. (eds.) Modeling,
  {Systems} {Engineering}, and {Project} {Management} for {Astronomy} {VIII}.
  p.~29. SPIE, Austin, United States (Jul 2018). \doi{10.1117/12.2310184}

\bibitem{kontio2008focus}
Kontio, J., Bragge, J., Lehtola, L.: The focus group method as an empirical
  tool in software engineering. In: Guide to advanced empirical software
  engineering, pp. 93--116. Springer (2008)

\bibitem{lambourne_2017}
Lambourne, J.: Why we use a 'docs as code' approach for technical documentation
  (Aug 2017),
  \url{https://technology.blog.gov.uk/2017/08/25/why-we-use-a-docs-as-code-approach-for-technical-documentation/}

\bibitem{louadah_towards_2014}
Louadah, H., Champagne, R., Labiche, Y.: Towards automating {Interface}
  {Control} {Documents} elaboration and management. vol.~1250, pp. 26--33
  (2014)

\bibitem{maier1998architecting}
Maier, M.W.: Architecting principles for systems-of-systems. Systems
  Engineering: The Journal of the International Council on Systems Engineering
  \textbf{1}(4),  267--284 (1998)

\bibitem{ozerova2020comparison}
Ozerova, M.I., Zhigalov, I.E., Vershinin, V.V.: Comparison of document
  generation algorithms using the docs-as-code approach and using a text
  editor. In: Proceedings of the Computational Methods in Systems and Software.
  pp. 315--326. Springer (2020)

\bibitem{petersen2014action}
Petersen, K., Gencel, C., Asghari, N., Baca, D., Betz, S.: Action research as a
  model for industry-academia collaboration in the software engineering
  context. In: Proceedings of the 2014 international workshop on Long-term
  industrial collaboration on software engineering. pp. 55--62 (2014)

\bibitem{rahmani_managing_2011}
Rahmani, K., Thomson, V.: Managing subsystem interfaces of complex products.
  International Journal of Product Lifecycle Management  \textbf{5}(1), ~73
  (2011). \doi{10.1504/IJPLM.2011.038103}

\bibitem{rong2020devdocops}
Rong, G., Jin, Z., Zhang, H., Zhang, Y., Ye, W., Shao, D.: Devdocops: Enabling
  continuous documentation in alignment with devops. Software: Practice and
  Experience  \textbf{50}(3),  210--226 (2020)

\bibitem{sheard2018incose}
Sheard, S., Creel, R., Cadigan, J., Marvin, J., Chim, L., Pafford, M.E.: Incose
  working group addresses system and software interfaces. INSIGHT
  \textbf{21}(3),  62--71 (2018)

\bibitem{thomchick2018improving}
Thomchick, R.: Improving access to api documentation for developers with
  docs-as-code-as-a-service. Proceedings of the Association for Information
  Science and Technology  \textbf{55}(1),  908--910 (2018)

\bibitem{tsui2018digital}
Tsui, R., Davis, D., Sahlin, J.: Digital engineering models of complex systems
  using model-based systems engineering (mbse) from enterprise architecture
  (ea) to systems of systems (sos) architectures \& systems development life
  cycle (sdlc). In: INCOSE International Symposium. vol.~28, pp. 760--776.
  Wiley Online Library (2018)

\bibitem{vipavetz_interface_2016}
Vipavetz, K., Shull, T.A., Infeld, S., Price, J.: Interface {Management} for a
  {NASA} {Flight} {Project} using {Model}-{Based} {Systems} {Engineering}
  ({MBSE}). INCOSE International Symposium  \textbf{26}(1),  1129--1144 (2016)

\bibitem{wheatcraft20109}
Wheatcraft, L.S.: 9.2. 2 everything you wanted to know about interfaces, but
  were afraid to ask. In: INCOSE International Symposium. vol.~20, pp.
  1132--1149. Wiley Online Library (2010)

\bibitem{wieringa2014single}
Wieringa, R.J.: Single-case mechanism experiments. In: Design Science
  Methodology for Information Systems and Software Engineering, pp. 247--267.
  Springer (2014)

\bibitem{wieringa2014technical}
Wieringa, R.J.: Technical action research. In: Design science methodology for
  information systems and software engineering, pp. 269--293. Springer (2014)

\bibitem{yasseri_interface_2019}
Yasseri, S.F., Bahai, H.: Interface and integration management for {FPSOs}.
  Ocean Engineering  \textbf{191},  106441 (Nov 2019)

\bibitem{zdravkovic2017challenges}
Zdravkovi{\'c}, M., Panetto, H.: The challenges of model-based systems
  engineering for the next generation enterprise information systems (2017)

\end{thebibliography}

\end{document}